\documentclass[12pt]{article}
\begin{document}
\renewcommand{\theequation}{\thesection.\arabic{equation}}
\newcommand{\eqn}[1]{(\ref{#1})}
\newcommand{\pp}{{=\!\!\!|}}
\renewcommand{\section}[1]{\addtocounter{section}{1}
\vspace{5mm} \par \noindent
  {\bf \thesection . #1}\setcounter{subsection}{0}
  \par
   \vspace{2mm} } 
\newcommand{\sectionsub}[1]{\addtocounter{section}{1}
\vspace{5mm} \par \noindent
  {\bf \thesection . #1}\setcounter{subsection}{0}\par}
\renewcommand{\subsection}[1]{\addtocounter{subsection}{1}
\vspace{2.5mm}\par\noindent {\em \thesubsection . #1}\par
 \vspace{0.5mm} }
\renewcommand{\thebibliography}[1]{ {\vspace{5mm}\par \noindent{\bf
References}\par \vspace{2mm}}
\list
 {\arabic{enumi}.}{\settowidth\labelwidth{[#1]}\leftmargin\labelwidth
 \advance\leftmargin\labelsep\addtolength{\topsep}{-4em}
 \usecounter{enumi}}
 \def\newblock{\hskip .11em plus .33em minus .07em}
 \sloppy\clubpenalty4000\widowpenalty4000
 \sfcode`\.=1000\relax \setlength{\itemsep}{-0.4em} }
\def\a{\hat{a}}
\def\ba{\hat{\bar a}}
\def\b{\hat{b}}
\def\bb{\hat{\bar b}}
\def\hd{\hat{D}}
\def\cd{\check{D}}
\def\hq{\hat{Q}}
\def\hm{\widehat{M}}
\def\hn{\widehat{N}}
\def\p{{\dot{+}}}
\def\m{{\dot{-}}}
\def\ss{\scriptstyle}
\newcommand\rf[1]{(\ref{#1})}
\def\nn{\nonumber}
\newcommand{\sect}[1]{\setcounter{equation}{0} \section{#1}}
\renewcommand{\theequation}{\thesection .\arabic{equation}}
\newcommand{\NPB}[3]{{Nucl.\ Phys.} {\bf B#1} (#2) #3}
\newcommand{\CMP}[3]{{Commun.\ Math.\ Phys.} {\bf #1} (#2) #3}
\newcommand{\PRD}[3]{{Phys.\ Rev.} {\bf D#1} (#2) #3}
\newcommand{\PLB}[3]{{Phys.\ Lett.} {\bf B#1} (#2) #3}
\newcommand{\JHEP}[3]{{JHEP} {\bf #1} (#2) #3}
\newcommand{\ft}[2]{{\textstyle\frac{#1}{#2}}}
\def\st{\scriptstyle}
\def\sst{\scriptscriptstyle}
\def\mco{\multicolumn}
\newcommand\text[1]{\rm #1}
\newcommand\WL{{dWL}}
\newcommand\order[1]{\vert_{\theta^#1}}
\newcommand\Border[1]{\Big\vert_{\theta^#1}}

\thispagestyle{empty}

\begin{center}

\hfill VUB/TENA/99/03\\ [3mm]
\hfill{\tt hep-th/9905141}\\
 
\vspace{3cm}

{\large\bf Properties of Semi-Chiral Superfields}\\

\vspace{1.4cm}

{\sc J. Bogaerts\footnote{Aspirant FWO}, 
A. Sevrin, S. van der Loo and S. Van Gils }\\
             
\vspace{1.3cm}

{\em Theoretische Natuurkunde, Vrije Universiteit Brussel} \\
{\em Pleinlaan 2, B-1050 Brussel, Belgium} \\
{\footnotesize \tt bogaerts, 
asevrin, saskia@tena4.vub.ac.be}\\

\vspace{1.2cm}

\centerline{\bf Abstract}
\vspace{- 4 mm}  \end{center}
\begin{quote}\small
Whenever the $N=(2,2)$ supersymmetry algebra of non-linear $\sigma$-models in two 
dimensions does not close off-shell, a holomorphic two-form can be 
defined. The only known superfields providing candidate auxiliary fields 
to achieve an off-shell formulation are semi-chiral fields. Such a semi-chiral 
description is only possible when the two-form is constant. Using an 
explicit example, hyper-K\"ahler manifolds, we show that this is not 
always the case. Finally, we give a concrete construction of semi-chiral 
potentials for a class of hyper-K\"ahler manifolds using the duality exchanging a pair 
consisting of a chiral and a twisted-chiral superfield for one semi-chiral 
multiplet. 
\end{quote}
\vfill
\leftline{\sc May 1999}
\newpage
\baselineskip18pt
\addtocounter{section}{1}
\par \noindent
{\bf \thesection . Introduction}
  \par
   \vspace{2mm} 
\setcounter{footnote}{0}
\noindent  Starting with Zumino's discovery that the scalar fields of an $N=1$ non-linear 
$\sigma$-model in four dimensions should be viewed as coordinates on a 
K\"ahler manifold \cite{bruno}, ample evidence for the interplay between 
supersymmetry and complex geometry was found. 

Requiring more supersymmetry 
or raising the dimension puts further restrictions on the geometry, while lowering the 
dimensions relaxes the requirements.  Examples of the former statement are well known in 4 dimensions
where passing from $N=1$ to $N=2$ supersymmetry restricts 
the geometry of the scalars in vector multiplets to 
so-called special K\"ahler manifolds \cite{sk}. Similarly, the scalars in 
hyper-multiplets describe special hyper-K\"ahler manifolds \cite{shk}.

Particularly interesting examples of 
supersymmetric models in lower dimensions are two-dimensional non-linear $\sigma$-models
which are used e.g. for the world-sheet description of stringtheory. 
The closest analog in two dimensions of $N=1$ supersymmetry in four dimensions is 
$N=(2,2)$ supersymmetry. As long as torsion is absent, the target manifold 
is indeed K\"ahler \cite{luis} and an off-shell description is known. 
However, once torsion is present,
the geometry becomes much richer \cite{GHR}, \cite{other}
and an off-shell description is much harder to achieve. Finding an 
off-shell description has been the subject of numerous studies, 
\cite{GHR}, \cite{BLR}, \cite{RSS}, \cite{martinnew} and 
\cite{french}, which culminated in \cite{icproc}. There strong evidence 
was put forward to support the conjecture that chiral, twisted-chiral and 
semi-chiral superfields are sufficient to give a manifest off-shell 
description of these models. These are the only superfields which can be 
defined by constraints on a general $N=(2,2)$ superfield which are linear 
in the fermionic derivatives. Several explicit examples are known. All 
K\"ahler-manifolds can be described by chiral superfields, the $SU(2)\times U(1)$ 
Wess-zumino-Witten model is described either by a chiral and a twisted-chiral field
\cite{RSS} or by a semi-chiral multiplet \cite{icproc}. Finally, the 
$SU(2)\times SU(2)$ Wess-Zumino-Witten model requires one semi-chiral and 
one chiral multiplet \cite{icproc}. Various models dual to the above 
mentioned models were constructed in \cite{martinnew}, \cite{RSS}, \cite{neuc}.

In the present paper we focus on the case where the $N=(2,2)$ does not 
close off-shell in all directions and investigate under which conditions 
semi-chiral superfields provide an off-shell formulation. 
Non-closure implies the existence of a holomorphic two-form. 
A necessary condition for the semi-chiral description to be possible is 
that there exists a complex coordinate system in which this two-form
is constant. Using a 
particular example, hyper-K\"ahler manifolds, we are able to show that 
semi-chiral fields alone are not able to give a full off-shell description
thus falsifying the conjecture given in \cite{icproc}.
Semi-chiral potentials do describe hyper-K\"ahler manifolds provided the 
potential satisfies a non-linear differential equation. 

We end the paper
with a short study of duality transformations involving semi-chiral 
fields. An interesting duality brings a model described by one chiral and 
one twisted-chiral superfield to a model formulated in terms of one 
semi-chiral multiplet \cite{neuc}, \cite{martinnew}. If the original model has an $N=(4,4)$ 
supersymmetry, which is true when the potential satisfies the Laplace 
equation \cite{GHR}, then the dual model is a hyper-K\"ahler manifold. 
In this way we generate a class of solutions to the non-linear differential
equation using solutions of a linear differential equation.
This construction is similar in spirit to the construction in \cite{hk1}, 
where hyper-K\"ahler potentials were constructed using the duality between 
a real linear and a chiral superfield. 
\setcounter{equation}{0}
\section{$N=(2,2)$ supersymmetric non-linear $\sigma$-models}
\noindent A bosonic non-linear $\sigma$-model in two dimensions is characterized by 
a manifold, the target manifold, endowed with a metric $G_{\mu \nu}$ and a closed 
3-form $T_{\mu \nu\rho}$. 
Locally, the torsion can be written as the exterior derivative of the torsion potential $B_{\mu \nu}$, 
\begin{eqnarray}
T_{\mu \nu\rho}=-\frac 3 2 \partial_{[\mu }B_{\nu\rho]}
\end{eqnarray}
Such a model can be promoted to an $N=(1,1)$ supersymmetric model without any 
additional conditions on the geometry. However, passing from $N=(1,1)$ to $N=(2,2)$ 
supersymmetry requires 
further structure. Two (1,1) tensors $J_+^\mu {}_\nu$ and  $J_-^\mu 
{}_\nu$ are needed which satisfy
\begin{eqnarray}
J_\pm^\mu {}_\rho J_\pm^\rho{}_\nu=-\delta^\mu{}_\nu,\label{c1}\\
N[J_\pm]^\mu {}_{\nu\rho}\equiv J_\pm^\sigma{}_{[\nu}J_\pm^\mu {}_{\rho],\sigma}+
J_\pm^\mu {}_\sigma J_\pm^\sigma{}_{[\nu,\rho]}=0,\label{c2}\\
J_\pm^\rho{}_\mu J_\pm^\sigma{}_\nu G_{\rho\sigma}=G_{\mu \nu}\label{c3},\\
\nabla^\pm_\rho J_\pm^\mu {}_\nu=0\label{c4},
\end{eqnarray}
where $\nabla^+$ and $\nabla^-$ denote covariant differentiation\footnote{Covariant 
derivatives are taken as $\nabla_\nu V^\mu \equiv
V^\mu {}_{,\nu}+\Gamma^{\mu }{}_{\rho\nu}V^\rho$ and $\nabla_\nu V_\mu \equiv
V_{\mu ,\nu}-\Gamma^{\rho}{}_{\mu \nu}V_\rho$.} using the 
$\Gamma^\mu_{+\nu\rho}\equiv\left\{ {}^{\, \mu }_{\nu\rho}\right\}+T^\mu{}_{\nu\rho}$ 
and
$\Gamma^\mu_{-\nu\rho}\equiv\left\{ {}^{\, \mu }_{\nu\rho}\right\}-T^\mu{}_{\nu\rho}$
connections resp. 
The first two conditions arise from requiring that the supersymmetry 
algebra is satisfied {\it on-shell} and the last two conditions follow 
from the invariance of the action. Eqs. (\ref{c1}) and (\ref{c2}) imply that 
both $J_+$ and $J_-$ are complex structures. Eq. (\ref{c3}) imposes hermiticity of 
the metric with respect to both complex structures and eq. (\ref{c4}) states that 
both complex structures are covariantly constant but, when torsion is 
present, with respect to different connections.

The $N=(2,2)$ models characterized by eqs. (\ref{c1}-\ref{c4}) realize the $N=(2,2)$ supersymmetry
algebra {\it on-shell} only. One can show that the off-shell 
non-closing terms in the algebra, are proportional to the commutator of 
the two complex structures, $[J_+,J_-]$ \cite{icproc}. The construction of a manifest 
off-shell supersymmetric version of these model was the subject of intense 
investigations \cite{GHR}, \cite{BLR}, \cite{RSS}, \cite{martinnew}, \cite{icproc}.
Locally, the cotangent space can be decomposed as
\begin{eqnarray}
\ker [J_+,J_-]\oplus (\ker [J_+,J_-])^\perp=
\ker (J_+-J_-)\oplus \ker (J_++J_-)\oplus(\ker [J_+,J_-])^\perp.
\end{eqnarray}
In \cite{martinnew}, it was shown that  $\ker (J_+-J_-)$ and $\ker (J_++J_-)$ 
are integrable to chiral and twisted-chiral superfields resp. These 
superfields count as many components as $N=(1,1)$ superfields. Indeed, as 
the algebra closes off-shell in these directions, one does not expect that 
any new auxiliary fields are needed. 

Chiral and twisted chiral superfields
separately describe K\"ahler manifolds. However when both of them are 
simultanously present, the resulting manifold exhibits a product structure 
which projects on two K\"ahler subspaces \cite{GHR}. The complete manifold is not 
K\"ahler, which can be seen from the fact that it has torsion. 

In \cite{icproc}, it was shown that the dimension of 
$(\ker [J_+,J_-])^\perp$ is a multiple of four (see also further in this section). 
Furthermore, 
off-shell closure of the algebra requires additional auxiliary fields 
compared to the manifestly $N=(1,1)$ formulation of the model.  Only one 
class of superfields defined by constraints linear in the derivatives  
satisfies these requirements: the semi-chiral superfields\cite{BLR}. This, 
combined with several non-trivial  examples, led to the conjecture 
\cite{icproc} that semi-chiral superfields are sufficient to describe
$(\ker [J_+,J_-])^\perp$. In the present paper, we will give a class of 
explicit examples disproving the conjecture. 

From now on we focus our attention on the case where $\ker [J_+,J_-] =\emptyset$. 
Denoting ${\cal C}\equiv [J_+,J_-]$, we construct a non-degenerate two-form 
$\omega(U,V)\equiv G(U,{\cal C}^{-1}V)$ or in local coordinates
\begin{eqnarray}
\omega_{\mu \nu}=G_{\mu \rho}({\cal C}^{-1}){}^\rho{}_\nu.
\end{eqnarray}
The inverse of the commutator 
${\cal C}^{-1}$ can be written as the formal power series
\begin{eqnarray}
{\cal C}^{-1}=\sum_{n\geq 0}(J_-J_+)^{2n+1}=
-\sum_{n\geq 0}(J_+J_-)^{2n+1}.
\end{eqnarray}
The two-form satisfies $\omega(J_\pm U,J_\pm V)=-\omega(U,V)$.

Introducing complex coordinates $z^\alpha $ and $\bar z^{\bar\alpha 
}=(z^\alpha )^*$, we diagonalize $J_+$: $J^\alpha_{+\beta}= i\delta^\alpha 
_\beta$ and $J^\alpha _{+\bar\beta}=0$. 
In complex coordinates, only $\omega_{\alpha \beta}$ and its complex 
conjugate are non-vanishing.
Eqs. (\ref{c2}) and (\ref{c3}) for $J_+$ imply 
that $T_{\alpha \beta\gamma}=0$ and $G_{\alpha \beta}=0$ resp. Finally eq. 
(\ref{c4}) for $J_+$ yields $\Gamma_{+\alpha \beta \bar\gamma}= \Gamma_{- 
\alpha \bar\gamma\beta}=0$. Combining this with eq. (\ref{c4}) for $J_-$ gives
\begin{eqnarray}
\partial_{\bar\alpha }\omega_{\beta\gamma}=0.
\end{eqnarray}

Finally, let us give a very short proof that $\ker[J_+,J_-]=\emptyset$ 
implies that $d=4n$ with $n\in {\bf N}$. We view $\omega_{\alpha \beta}$ as the components
of an anti-symmetric $d/2\times d/2$ matrix. From  $\ker[J_+,J_-]=\emptyset$
combined with the non-degeneracy of the metric, we get that its 
determinant is non-vanishing implying that $d/2$ should be even. 

\setcounter{equation}{0}
\section{Semi-chiral parametrization}
\noindent  
We denote the semi-chiral coordinates by $r^a$, $\bar r^{\bar a}$, 
$s^{\hat a}$ and $s^{\hat{\bar a}}$, $a$, $\bar a$, $\a$, $\ba$ $\in 
\{1,\cdots n\}$, and we introduce a real function $K(r,\bar r, s, \bar s)$,
the semi-chiral potential. 
It is determined modulo the transformation $K\propto K+f(r)+\bar f (\bar r)+
g(s)+\bar g (\bar s)$ with $f(r)$ and $g(s)$ arbitrary holomorphic functions of $r$ and $s$
resp.
The potential is the Lagrange density in 
$N=(2,2)$ superspace. Passing to $N=(1,1)$ superspace and eliminating the 
auxiliary fields through their equations of motion yields explicit 
expressions for the metric, torsion potential and the complex structures 
\cite{icproc}.

In order to facilitate the notation, we introduce 
the $2n\times 2n$ matrices $L$, $N$, $M$, $\hn$ and $\hm$.
\begin{eqnarray}
L\equiv \left( \begin{array}{cc} K_{\a b} & K_{\a\bar b}\\
K_{\ba b} & K_{\ba\bar b}\end{array}\right), \qquad
N\equiv \left( \begin{array}{cc} K_{ab} & K_{ a{\bar b}}\\
K_{{\bar a} b} & K_{\bar a\bar b}\end{array}\right),\qquad
\hn\equiv \left( \begin{array}{cc} K_{\a\b} & K_{\a\bb}\\
K_{\ba\b} & K_{\ba\bb}\end{array}\right),
\label{defmat1}
\end{eqnarray}
\begin{eqnarray}
M\equiv \left( \begin{array}{cc} 0 & K_{ a{\bar b}}\\
K_{{\bar a} b} & 0\end{array}\right),\qquad
\hm\equiv \left( \begin{array}{cc} 0 & K_{\a\bb}\\
K_{\ba\b} & 0\end{array}\right),
\label{defmat2}
\end{eqnarray}
where e.g. $K_{a\bar b}$ stands for $K_{a\bar b}\equiv
\frac{\partial^2K}{\partial r^a\partial {\bar r}^{\bar b}}$.
Finally we also need the matrix $P$, defined by 
\begin{eqnarray}
P\equiv \left( \begin{array}{cc} {\bf 1} & 0\\
0 & -{\bf 1}\end{array}\right). 
\label{defmat3}
\end{eqnarray}
In terms of these matrices, the complex structures are given 
by\footnote{Rows and columns are labeled as $r\bar rs\bar s$ and we rescaled
$J_-$ by a factor $-1$.}
\begin{eqnarray}
J_+&=&\left(\begin{array}{cc}iP&0\\
-2iL^{-1}{}^T M P & i L^{-1}{}^T
P  L^T 
\end{array}\right),\nonumber\\
J_- &=&\left(\begin{array}{cc}iL^{-1}PL& 2iL^{-1} P \hm \\
0&iP \end{array}\right).\label{cs1}
\end{eqnarray}
The metric and torsion potential have simple expressions in terms of the 
complex structures,
\begin{eqnarray}
G&=& \frac 1 2 
\left(\begin{array}{cc}0& +L^T \\
-L&0 \end{array}\right)
[J_+,J_-]\label{met0}\\
B&=& \frac 1 2 
\left(\begin{array}{cc}0& +L^T \\
-L&0 \end{array}\right)
\{J_+,J_-\}.   \label{met1}
\end{eqnarray}
Eq. (\ref{met0}) clearly shows that the vanishing of $\ker [J_+,J_-]$ 
is necessary and sufficient for the existence of a non-degenerate metric.
Furthermore, the potential should be such that $\det L\neq 0$. Eq. 
(\ref{met0}) gives the explicit form for the two-form $\omega$,
\begin{eqnarray}
\omega_{a\b}=\frac 1 2 K_{a\b},\qquad \omega_{a\bb}=\frac 1 2 K_{a\bb}.
\end{eqnarray}

Quite remarkable is the existence of simple coordinate transformations 
which diagonalize either $J_+$ or $J_-$. Consider
\begin{eqnarray}
r^a&\rightarrow& z^a=r^a\nonumber\\
s^{\a}&\rightarrow& w_{\a}=K_a, \label{transfo}
\end{eqnarray}
then $J_+\rightarrow J'_+$ with
\begin{eqnarray}
J'_+= \left( \begin{array}{cc} iP & 0\\
0 & iP\end{array}\right), 
\end{eqnarray}
and $G\rightarrow G'$ and $B\rightarrow B'$ where
\begin{eqnarray}
G'&=& \frac 1 2 
\left(\begin{array}{cc}0& +{\bf 1} \\
-{\bf 1}&0 \end{array}\right)
[J'_+,J'_-]\nonumber\\
B'&=& \frac 1 2 
\left(\begin{array}{cc}0& +{\bf 1} \\
-{\bf 1}&0 \end{array}\right)
\{J'_+,J'_-\}.   \label{met2}
\end{eqnarray}
Rows and columns are labeled as $z$, $\bar z$, $w$ and $\bar w$.
Note that there is also a simple coordinate transformation which diagonalizes 
$J_-$ which is obtained by reversing the roles of $r$ and $s$ in the 
previous expressions. 

Eq. (\ref{met2}) now gives the two-form $\omega$ in complex coordinates,
\begin{eqnarray}
\omega_{a}{}^{\b}=\frac 1 2 \delta_a^b.
\end{eqnarray}
So we reach the conclusion that {\it a necessary condition for a
semi-chiral parametrization to be possible is the existence of a
complex coordinate system in which the 
two-form $\omega$ is constant}! Note that for $d=4$, we can always find a 
holomorphic coordinate transformation which makes the two-form $\omega$ 
constant. For $d>4$ this is not the case anymore.

Finally, for the sake of completeness, let us mention that these models 
are 1-loop UV finite provided they are Ricci flat where the Ricci 
tensor is computed using the connection with torsion. In \cite{marc} the 
one-loop $\beta$-function was directly computed in $N=(2,2)$ superspace. 
Its vanishing yields a constraint on the potential which requires the 
existence of a holomorphic function of $r$, $f(r)$ and a holomorphic 
function of $s$, $g(s)$, such that
\begin{eqnarray}
\frac{\det {\cal N}_2}{\det  {\cal N}_1}=(-1)^n|f(r)|^2|g(s)|^2, \label{uv}
\end{eqnarray}
where $n$ is the number of semi-chiral multiplets and ${\cal N}_1$ and ${\cal N}_2$ are 
$2n\times 2n$ matrices given by
\begin{eqnarray}
{\cal N}_1\equiv \left( \begin{array}{cc} K_{\a b} &
K_{\a\bb}\\
K_{\bar a b} & K_{\bar a\bb}\end{array}\right),
\qquad
{\cal N}_2\equiv \left( \begin{array}{cc} K_{\a\bar b} &
K_{\a\bb}\\
K_{a \bar b} & K_{a\bb }\end{array}\right).\label{ndef}
\end{eqnarray}

\setcounter{equation}{0}
\section{Hyper-K\"ahler manifolds}
\noindent
As is clear from eq. (\ref{met0}), the semi-chiral parametrization is well 
defined, provided $\ker [J_+,J_-]=\emptyset$. The most familiar class of 
complex manifolds satisfying this are the hyper-K\"ahler manifolds. 
A hyper-K\"ahler manifold has three complex structures $J_i$, 
$i\in\{1,2,3\}$ which satisfy 
\begin{eqnarray}
J_iJ_j=-\delta_{ij}{\bf 1}+\varepsilon_{ijk}J_k,
\end{eqnarray}
and which are such that the manifold is K\"ahler with respect to all three of them. 
It is easy to see that a hyper-K\"ahler manifold has a two-sphere worth of 
complex structures. Indeed $\vec x\cdot \vec{J}=
\sum_{i=1}^3 x_i J_i$ is a complex structure 
provided that $\vec x\cdot\vec x=1$. 
Choosing e.g. $J_+=J_1$, and requiring that $\ker [J_+,J_-]=\emptyset$,
we find that $J_-$ 
can be any element of the two-sphere,
except for the north- and southpole, $\vec x = (\pm 1,0,0)$. 
Choosing for $J_-$ the north-pole, 
we obtain the description of the manifold in terms of chiral superfields, 
while choosing the south-pole we get the parametrization in terms of 
twisted-chiral superfields. Clearly, these are the only two choices where $J_+$ and 
$J_-$ commute. In other words, 
there is a cylinder worth of choices for $J_-$ where the algebra does not close off-shell
and which can potentially be described
by semi-chiral coordinates. In order for this to work, we need at least that the torsion
vanishes, $T=0$. Indeed, choosing one 
element of the cylinder $J_-=\vec x\cdot \vec J$, we get $\{J_+,J_-\}=-2x_1{\bf 
1}$. From eq. (\ref{met1}), one obtains that for a generic element of the cylinder, the 
torsion potential differs from zero, but the torsion still vanishes, $T=dB=0$.
Turning to the two-form $\omega$ introduced in section 2, one easily shows 
that, for the present choice for $J_+=J_1$ and  $J_-=\vec x\cdot \vec J$, 
it is given by
\begin{eqnarray}
\omega=\frac{1}{2(x_2^2+x_3^2)}(x_3\omega_2-x_2\omega_3),
\end{eqnarray}
where $\omega_i$ are the fundamental two-forms of the hyper-K\"ahler 
manifold, defined by $\omega_i(U,V)\equiv G(U,J_iV)$, $i\in\{1,2,3\}$. As was shown in the 
previous section, a semi-chiral parametrization is only possible, if 
complex coordinates exist, diagonalizing $J_+$, where $\omega$ is 
constant. In this case $\omega$ is given as a linear combination of the 
the two fundamental two-forms $\omega_2$ and $\omega_3$. Several explicit 
examples are known of higher dimensional hyper-K\"ahler manifolds where this
is not the case \cite{hk1}.

As was already mentioned, there is still the simplest case $d=4$, where 
both $\omega_2$ and $\omega_3$ can be made constant through a holomorphic 
coordinate transformation. However we find, as we will see in next section,
that the metric satisfies the Monge-Ampere equation after passing from 
semi-chiral to complex coordinates. To achieve this on a hyper-K\"ahler 
manifold, one already needs a holomorphic coordinate transformation. The 
residual holomorphic coordinate transformations can now only turn the 
two-form to a constant provided the two-form was originally a phase, which as 
far as we know, is not necessarily true. Nonetheless, many interesting examples are of 
this kind. In particular, the four dimensional hyper-K\"ahler manifolds 
constructed in \cite{hk1} are of this form. This includes familiar 
examples such as the multi-Eguchi-Hanson \cite{eh} and Taub-NUT \cite{hawgib} self-dual 
instantons.

\setcounter{equation}{0}
\section{The four-dimensional case}
\noindent
The simplest hyper-K\"ahler manifolds are the four dimensional ones. We 
will choose $J_+=J_1$ and $J_-=J_2$. Requiring $\{J_+,J_-\}=0$ we find 
using eq. (\ref{cs1}) that the semi-chiral potential should satisfy
\begin{eqnarray}
|K_{rs}|^2+|K_{r\bar s}|^2=2K_{r\bar r}K_{s\bar s}. \label{cac1}
\end{eqnarray}
Performing the coordinate transformation eq. (\ref{transfo}), we find that 
the metric eq. (\ref{met2}) satisfies the 
Monge-Amp\`ere equation,
\begin{eqnarray}
G'_{z\bar z}G'_{w\bar w}- G'_{z\bar w} G'_{w\bar z}=1,\label{moam}
\end{eqnarray}
iff. eq. (\ref{cac1}) holds. 
Indeed, in four dimensions one finds that eq. (\ref{met0}) yields the 
following non-vanishing components of the metric (and their complex conjugates),
\begin{eqnarray}
&&G_{rr}=4\Omega^{-1}K_{r\bar r}K_{rs}K_{r\bar s},\quad
G_{r\bar r}= 2\Omega^{-1}K_{r\bar r}\left( |K_{rs}|^2+|K_{r\bar 
s}|^2\right),\quad\nonumber\\
&&G_{rs}= 2\Omega^{-1}K_{rs}\left( K_{r\bar r }K_{s\bar s}+|K_{r\bar 
s}|^2\right),\quad
G_{r\bar s}= 2\Omega^{-1}K_{r\bar s}\left( K_{r\bar r }K_{s\bar s}+|K_{r 
s}|^2\right),\nonumber\\
&&G_{ss}=4\Omega^{-1}K_{s\bar s}K_{rs}K_{\bar r s},\quad
G_{s\bar s}= 2\Omega^{-1}K_{s\bar s}\left( |K_{rs}|^2+|K_{r\bar 
s}|^2\right),
\end{eqnarray}
where $\Omega\equiv |K_{rs}|^2-|K_{r\bar s}|^2$.  After the coordinate 
transformation eq. (\ref{transfo}), the components of the metric are given 
by
\begin{eqnarray}
G'_{w\bar w}=2\Omega^{-1}K_{s\bar s},\quad
G'_{z\bar w}=2\Omega^{-1}\left(K_{rs} K_{r\bar s}
-K_{rr}K_{s\bar s} \right),\nonumber\\
G'_{z\bar z}=G_{w\bar w}^{-1}\left(|G_{z\bar w}|^2+1\right)-
\left(2\Omega K_{s\bar s}\right)^{-1}(|K_{rs}|^2+|K_{r\bar s}|^2-
2K_{r\bar r}K_{s\bar s})^2.\label{d4}
\end{eqnarray}
As a result we get that eq. (\ref{cac1}) is 
indeed a necessary and sufficient condition on the potential so that it 
describes a hyper-K\"ahler potential. Note that, as expected, eq. (\ref{uv}) 
is satisfied with $f(r)=g(s)=1$.

In \cite{hk1}, a powerful method was developed to construct solutions to 
eq. (\ref{moam}), the Legendre transformation construction. In the 
remainder of this section we will discuss various duality transformations which involve
semi-chiral fields and we will give a construction of 
solutions to eq. (\ref{cac1}) analogous to the Legendre transform construction in \cite{hk1}.

In \cite{neuc} (see also \cite{martinnew}) various duality transformations in
$N=(2,2)$ superspace were catalogued. The simplest ones 
are those which do not need any isometries. They arise by passing to a 
first order action in superspace. I.e. the superfield constraints are 
imposed through Lagrange multipliers. The best known example is the 
duality between chiral and complex linear superfields (for a recent account see e.g. 
\cite{mil}). Both describe 
K\"ahler geometry and the former gives the minimal description while the 
latter is a non-minimal description. The two potentials are related 
through a simple Legendre transformation.

Similar duality transformations exist for semi-chiral superfields. One can 
perform a Legendre transformation either with respect to $r$ or with respect to $s$ or 
with respect to both of them. Given a potential $K(r,\bar r, s,\bar s)$, we can 
construct three potentials
\begin{eqnarray}
K_1(r,\bar r, s,\bar s)&=&K(r',\bar r', s,\bar s)-r'r-\bar r'\bar 
r,\nonumber\\
K_2(r,\bar r, s,\bar s)&=&K(r,\bar r, s',\bar s')-s's-\bar s'\bar 
s,\nonumber\\
K_3(r,\bar r, s,\bar s)&=&K(r',\bar r', s',\bar s')-r'r-\bar r'\bar 
r-s's-\bar s'\bar s,
\end{eqnarray}
where in the first case $r=K_{r'}$, in the second case $s=K_{s'}$ and in the 
last case $r=K_{r'}$ and $s=K_{s'} $ hold. One verifies immediately that 
if $K$ satisfies eq. (\ref{cac1}) then so do $K_1$, $K_2$ and $K_3$. 
These three duality transformations simply shuffle around the auxiliary field content of
the system and act as mere coordinate transformations on the physical fields. 

In case isometries are present more interesting duality transformations become possible. 
The most typical semi-chiral example is the one which interchanges one 
semi-chiral multiplet for one chiral and one twisted chiral multiplet. 
The geometry obviously changes now. In the present case the semi-chiral 
coordinates describe a hyper-K\"ahler manifold, while the chiral/twisted 
chiral combination describes a manifold with a product structure which has 
e.g. a non-trivial torsion. At the chiral/twisted-chiral side the model 
shows a simple enhancement of the supersymmetry to $N=(4,4)$, provided the 
potential satisfies the Laplace equation. The dual potential turns out to 
describe a hyper-K\"ahler manifold. By this construction, one obtains 
immediately the semi-chiral parametrization of the hyper-K\"ahler 
manifold. The advantage of this construction is that the non-linear 
differential equation (\ref{cac1}) gets replaced by a linear differential 
equation, the Laplace equation.

The starting point is a real prepotential 
$F(x,u,\bar u)$, where $x\in{\bf R}$ and $u\in{\bf C}$, 
which satisfies the Laplace equation
\begin{eqnarray}
F_{xx}+F_{u\bar u}=0.\label{lt}
\end{eqnarray} 
This combines two requirements: the chiral/twisted-chiral potential, which is precisely
the prepotential under consideration, simultanously
exihibits an Abelian isometry and it has $N=(4,4)$ supersymmetry. Full 
details can be found in the appendix. The present coordinates $x$ and $u$ 
are related to the chiral and twisted-chiral coordinates $z$ and $w$ by
$x=-i(z-\bar z+ w-\bar w)$, $u=z+\bar w$ and $\bar u = u^*$.

The semi-chiral potential $K(r-\bar r,r+\bar s,\bar r+s)$ is obtained from 
the prepotential through a Legendre transformation,
\begin{eqnarray}
K(r-\bar r,r+\bar s,\bar r+s) &=&F(x,u,\bar u)-\frac u 2 (r+\bar r + 2\bar 
s) - \nonumber\\
&&\frac {\bar u}{ 2} (r+\bar r + 2 s) -\frac{ix}{2}(r-\bar r), \label{lt1}
\end{eqnarray}
where
\begin{eqnarray}
F_x&=&\frac i 2 (r-\bar r),\nonumber\\
F_u&=&\frac 1 2 (r+\bar r + 2 \bar s),\nonumber\\
F_{\bar u}&=&\frac 1 2 (r+\bar r + 2  s).\label{lt2}
\end{eqnarray}
Using eqs. (\ref{lt}), (\ref{lt1}) and (\ref{lt2}), we get
\begin{eqnarray}
K_{rs}&=& \frac {1}{ 2\Lambda} (F_{xu}^2-|F_{xu}|^2+iF_{xx}F_{xu}-F_{xx}^2+i 
F_{uu}F_{x\bar u}- F_{xx}F_{uu}),
\nonumber\\
K_{r\bar s}&=& \frac{ 1}{ 2\Lambda} (F_{x\bar u}^2-|F_{xu}|^2+iF_{xx}F_{x\bar u}-F_{xx}^2+i 
F_{\bar u\bar u}F_{x u}- F_{xx}F_{\bar u\bar u}),
\nonumber\\
K_{r\bar r}&=&
\frac{ 1}{ 4\Lambda} ((F_{xu}-F_{x\bar u})^2-|F_{uu}+F_{xx}|^2),
\nonumber\\
K_{s\bar s}&=&-\frac{1}{\Lambda}(F_{xx}^2+|F_{xu}|^2),  \label{d5}
\end{eqnarray}
where
\begin{eqnarray}
\Lambda=-F_{xx}^3+F_{xx}|F_{uu}|^2-2F_{xx}|F_{xu}|^2-F_{xu}^2F_{\bar u\bar 
u}-F_{x\bar u}^2F_{uu}. \label{lamm}
\end{eqnarray}
Using this, one immediately checks that the resulting semi-chiral potential 
satisfies eq. (\ref{cac1}). In other words, dualizing a 
chiral/twisted-chiral potential having an $N=(4,4)$ supersymmetry and an 
Abelian isometry yields a semi-chiral potential which describes a 
hyper-K\"ahler manifold! The resulting potential obviously still has an 
Abelian isometry, $\delta r =\varepsilon$ and $\delta s=-\varepsilon$ with 
$\varepsilon\in{\bf R}$. 

This construction is strikingly similar to the Legendre transform 
construction in \cite{hk1}, which follows from the duality between
an $N=(4,4)$ model described by a real linear and a chiral superfield and an 
$N=(4,4)$ model described by two chiral superfields. Again this duality requires an
Abelian isometry. There too, eq. (\ref{lt}) was the 
starting point for the construction of hyper-K\"ahler potentials directly 
in local complex coordinates. From a real prepotential $\hat F(x,v,\bar v)$, satisfying
$\hat F_{xx}+  \hat F_{v \bar v}=0$, one obtains the K\"ahler potential, $K_{Kahler}$, 
through the Legendre transformation
\begin{eqnarray}
K_{Kahler}(v,\bar v, z, \bar z) =\hat F(x,v,\bar v) -(z+\bar z)x,
\end{eqnarray}
where
\begin{eqnarray}
\hat F_x=z+\bar z.
\end{eqnarray}
This allows for explicit expressions for the metric in terms of the 
prepotential,
\begin{eqnarray}
G'_{z\bar z}=-\hat F_{xx}^{-1},\quad G'_{z\bar v}=\hat F_{x\bar v}\hat F_{xx}^{-1},\label{d6}
\end{eqnarray}
and $G'_{\bar z v}=G'_{z\bar v}{}^*$ and $G'_{v\bar v}$ follows from the Monge-Amp\`ere
equation. The extra fundamental two-forms are constant.

Comparing both constructions is possible if we pass from semi-chiral to 
complex coordinates. For simplicity, we use the coordinate transformation
which diagonalizes $J_-$,
\begin{eqnarray}
r\rightarrow \bar w=K_s,\qquad s\rightarrow \bar z =s,
\end{eqnarray}
and from eq. (\ref{lt1}), one gets the identification $w=-u$.
The metric is then given by eq. (\ref{d4}) in which $r$ and $s$ are 
interchanged. Combining this with eq. (\ref{d5}) yields the expression for 
the metric in complex coordinates,
\begin{eqnarray}
G'_{z\bar z}=i(F_{xu}-F_{x\bar u})^{-1},\quad 
G'_{z\bar w}= i(F_{uu} +F_{xx})(F_{xu}-F_{x\bar u})^{-1} ,\label{d7}
\end{eqnarray}
where once more we obtain $G'_{w\bar w}$ from the Monge-Amp\`ere equation.
Comparing eq. (\ref{d6}) to eq. (\ref{d7}) we get, after identifying 
$u=-iv$, a relation between $F(x,v,\bar v)$ and $\hat F(x,v,\bar v)$,
\begin{eqnarray}
\hat F_x = F_v+ F_{\bar v} + \alpha ,\label{ff}
\end{eqnarray}
with $\alpha $ a real constant. Given either $F$ or $\hat F$, this allows 
the construction of $\hat F$ and $F$ resp. The requirement that the 
resulting prepotential satisfies the Laplace equation fully determines the 
prepotential $\hat F$, once $F$ is given. However, given $\hat F$, $F$ is 
only determined modulo a function $G(v-\bar v, x)$ which satisfies 
$G_{v}+G_{\bar v}=0$ and $G_{xx}=G_{vv}$. This has no influence on both 
eqs. (\ref{ff}) and (\ref{lt}) but it might be needed in order to have a 
well-defined legendre transformation in eq. (\ref{lt2}) which is equivalent to 
requiring that $\Lambda$ in eq. (\ref{lamm}) does not vanish. 

We give a few examples where each time both the complex and the semi-chiral prepotential 
are given. Each time it is straightforward to check that both satisfy the Laplace equation
and that they are related by eq. (\ref{ff}). 

The four-dimensional
special hyper-K\"ahler manifolds \cite{shk} are descibed by
\begin{eqnarray}
\hat F&=&-\left(v \bar f(\bar v)+\bar v f(v)\right)+
\frac{x^2}{2}\left(\partial_v f(v) +\partial_{\bar v} \bar f(\bar v)\right),\nonumber\\
F&=&x\left(f(v)+\bar f(\bar v)\right)+\frac a 2 x^2+\frac a 2 \left( v-\bar v\right)^2,
\end{eqnarray}
where $f(v)$ is a holomorphic function of $v$ and $\bar f(\bar v)$ its 
complex conjugate and $a$ is an arbitrary real constant.

In \cite{hk1}, two different representations of flat space were given
\begin{eqnarray}
\hat F_1&=&v\bar v-\frac 1 2 x^2\nonumber\\
F_1&=&-\frac x 2 (v+\bar v)+\frac 1 2 x^2+\frac 1 2 \left( v-\bar v\right)^2,
\end{eqnarray}
and
\begin{eqnarray}
\hat F_2&=&r-x\ln (x+r)+\frac 1 2 x\ln(4 v\bar v),\nonumber\\
F_2&=&\frac 1 8 (v-\bar v+x)\ln (v-\bar v-x)-
\frac 1 8 (v-\bar v -x)\ln(v-\bar v +x) +\nonumber\\
&&\frac 1 4 (v+\bar v)\ln (4 v\bar v)-\frac 1 2 x\ln (v+\bar v+r) -\frac 1 2(v+\bar v)\ln 
(x+r)+\nonumber\\
&&\frac 1 4 (v-\bar v)\ln\left( \frac{v}{\bar v}\frac{2(v-\bar v)\bar v+x(x+r)}
{2(v-\bar v)v-x(x+r)}\right), 
\end{eqnarray}
where
\begin{eqnarray}
r\equiv \sqrt{x^2 + 4 v \bar v}.
\end{eqnarray}
Other prepotentials for hyper-K\"ahler manifolds can now be constructed by 
superimposing  $F_1$ and $F_2$ or equivalently $\hat F_1$ and $\hat F_2$.
In this way one obtains the multi-Eguchi-Hanson manifolds \cite{eh,hawgib,hk1} 
by superimposing $F_2$ or $\hat F_2$ about different points $\vec\rho_A$,
\begin{eqnarray}
F_{EH}=\sum_{A=1}^{m+1} F_2(\vec r -\vec\rho_a),
\end{eqnarray}
or the Taub-NUT manifolds \cite{hawgib,hk1} by adding an $F_1$ to this
\begin{eqnarray}
F_{TN}=F_1(\vec r)+\sum_{A=1}^{m+1} F_2(\vec r -\vec\rho_a),
\end{eqnarray}
and similar expressions where the complex prepotentials are replaced by 
semi-chiral prepotentials.

\newpage

\setcounter{equation}{0}
\section{Conclusions}
\noindent
In \cite{icproc}, it was conjectured that chiral, twisted-chiral and 
semi-chiral superfields are sufficient to give a full off-shell, manifest supersymmetric 
description of $N=(2,2)$ supersymmetric non-linear $\sigma$-models in two 
dimensions. The chiral and twisted-chiral superfields do give a complete description of the 
directions along which the supersymmetry closes \cite{martinnew} while the 
semi-chiral superfields were expected to introduce the necessary auxiliary 
fields for those directions where no off-shell closure was achieved.

In the present paper we showed that this is not true. Non-closure of the 
$N=(2,2)$ supersymmetry implies the existence of a holomorphic two-form. Moving 
from semi-chiral coordinates to complex coordinates, one gets that this two-form 
is constant. Hyper-K\"ahler manifolds provide particularly interesting 
examples. Choosing left- and right-complex structures to be anti-commuting, 
we do get full non-closure of the algebra. The above mentioned two-form is 
the fundamental two-form associated with the ``third'' complex structure 
which is the product of the left with the right complex structure. A necessary 
condition for the semi-chiral parametrization to be possible is that this 
fundamental two-form is a constant, which is not the case for an arbitrary 
hyper-K\"ahler manifold!

As a result, the problem of finding a manifest supersymmetric description 
of $N=(2,2)$ non-linear $\sigma$-models is once more open. Chiral, 
twisted-chiral and semi-chiral superfields exhaust the superfields which 
can be defined by constraints linear in the superderivatives. What remains are 
constraints which are higher order in the derivatives. These have not been 
systematically studied, but past experience shows that most often they 
give non-minimal descriptions of known super-multiplets. Finally, there is 
a last possibility which would certainly work but which involves harmonic 
superspace. A drawback of this approach is that it is extremely hard to 
extract explicit expressions for metric and torsion.

Finally, we presented a systematic way to construct $d=4$ hyper-K\"ahler 
manifolds starting from an intriguing duality transformation between an 
$N=(4,4)$ model described by one chiral and one twisted chiral superfield 
and a $d=4$ hyper-K\"ahler manifold. In particular, this implies that 
well-known hyper-K\"ahler manifolds such as multi-Eguchi-Hanson and 
Taub-NUT have a dual which is a complex manifold with a product structure. 
The consequences of this duality transformation and the relation, if any,
with the non-abelian T-duals of these models given in \cite{kostas}, certainly
merit further study. 

\vspace{.75cm}


\newpage

\noindent {\bf Acknowledgements}: We thank Martin Ro\v cek, Kostas Sfetsos 
and Jan Troost for useful discussions. This work was supported in part 
by the European Commission TMR programme ERBFMRX-CT96-0045 in 
which all authors are associated to K.U. Leuven. 
        
\vspace{5mm}

\renewcommand{\theequation}{A.\arabic{equation}}
\setcounter{equation}{0}
\par \noindent
  {\bf Appendix: $N=(2,2)$ superspace}
  \par
   \vspace{2mm} 
\noindent
In this appendix we summarize some properties of $N=(2,2)$ superspace
and superfields, together with some aspects of duality transformations.
The fermionic coordinates which parametrize the $N=(2,2)$ superspace are denoted by
$\theta^+$, $\theta^\p\equiv -(\theta^+)^\dagger$, $\theta^-$ and 
$\theta^\m\equiv -(\theta^-)^\dagger$ and the bosonic 
coordinates by $x^\pp$ and $x^=$. The fermionic derivatives $D_+$, $D_\p\equiv
(D_+)^\dagger$, $D_-$ and $D_\m\equiv (D_-)^\dagger$ satisfy
\begin{eqnarray}
\{D_+,D_\p\}=2i\partial_{\pp},\qquad
\{D_-,D_\m\}=2i\partial_=,
\end{eqnarray}
with all other (anti-)commutators between derivatives vanishing. 
The $N=(1,1)$ superderivatives are given by the real part of the $N=(2,2)$ fermionic 
derivatives,
\begin{eqnarray}
\hd_+\equiv \frac 1 2 (D_++D_\p),\qquad \hd_-\equiv \frac 1 2 (D_-+D_\m ),
\end{eqnarray}
while the extra supersymmetry generators are then proportional to the 
imaginary part of the $N=(2,2)$ superderivatives,
\begin{eqnarray}
\hq_+\equiv \frac i 2 (D_+-D_\p),\qquad \hq_-\equiv \frac i 2 (D_--D_\m ).
\end{eqnarray}

Consider a set of general $N=(2,2)$ superfields, $X^\mu $, 
$\mu \in\{1,\cdots,d\}$. The most general constraints linear in the 
derivatives are
\begin{eqnarray}
\hq_+ X^\mu =J^\mu _+{}_\nu (X) \hd_+ X^\nu, \qquad
\hq_- X^\mu =J^\mu _-{}_\nu(X) \hd_- X^\nu.\label{co1}
\end{eqnarray}
A detailed analysis of the integrability conditions following from 
eq. (\ref{co1}) yields that both $J_+$ and $J_-$ are complex structures 
which mutually commute \cite{icproc}. Through a suitable coordinate 
transformation they can be diagonalized resulting in two classes of 
superfields, chiral and twisted-chiral \cite{GHR} superfields. 
\begin{itemize}
\item Chiral superfields, $z$ and $\bar z\equiv z^\dagger$,\\
$D_\p z=D_\m z=0$ or $\hat Q_+z=+i \hat D_+z$ and $\hat Q_-z=+i \hat D_-z$.
\item Twisted chiral superfields, $w$ and $\bar w\equiv w^\dagger$,\\
$D_\p w=D_- w=0$ or $\hat Q_+w=+i \hat D_+w$ and $\hat Q_-w=-i \hat D_-w$.
\end{itemize}
On a chiral superfield $dz$, both $J_+$ and $J_-$ have eigenvalue $+i$. On
a twisted chiral superfield $dw$ , one finds that $J_+$ has eigenvalue 
$+i$ while $J_-$ has eigenvalue $-i$. 
Chiral and twisted chiral fields have the same number of components as a general $N=(1,1)$ 
superfield, consistent with the fact that the algebra closes in the 
directions where the complex structures commute. 
A weaker set of constraints is still possible where only one 
chirality gets constrained. A detailed analysis shows that, in order to get  
non-trivial dynamics, they should occur in pairs, the members of which 
have constraints of opposite chirality. This results in semi-chiral 
superfields \cite{BLR}.
\begin{itemize}
\item Semi-chiral superfields, $r$, $\bar r\equiv r^\dagger$, $s$ and
$\bar s\equiv s^\dagger$,\\
$D_\p r=D_- s=0$ or $\hat Q_+ r = +i \hat D_+ r$ and $\hat Q_- s= -i \hat D_-s$.
\end{itemize}
Semi-chiral superfields contain twice as many components as an $N=(1,1)$ 
superfield, however, half of them turn out to be auxiliary.
On $dr$, $J_+$ is diagonal with eigenvalue $+i$, while on $ds$, $J_-$ is 
diagonal with eigenvalue $+i$. The precise action of $J_-$ on $dr$ and 
$J_+$ on $ds$ is model dependent and can only be obained after elimination 
of the auxiliary fields.

Other constraints, linear in the derivatives, are still possible, but they imply 
restrictions on the dependence of the superfields on the bosonic 
coordinates (see e.g. \cite{hub}). 
We do not consider this here as it does not seem relevant to 
the present case. 

Finally, we comment on duality transformations involving semi-chiral superfields.
It is well known that in the presence of an abelian isometry a chiral 
field can be dualized to a twisted-chiral superfield and vice-versa. 
Similarly \cite{martinnew,icproc}, when a specific abelian isometry is 
present, a pair consisting of a chiral and a twisted-chiral superfield can 
be dualized to a semi-chiral field and vice-versa. Consider
a model described by a chiral/twisted chiral potential $F(z,\bar z,w,\bar w)$, which
is such that an abelian isometry exists,
\begin{eqnarray}
\delta z=\varepsilon,\quad \delta w =-\varepsilon,  \label{iso}
\end{eqnarray}
with $\varepsilon$ a real constant. We introduce 
prepotentials $V$ and $W$ which are complex unconstrained superfields and
one set of semi-chiral superfields $r$ and $s$. We
consider the first order action
\begin{eqnarray}
{\cal S}&=&\int d^2xd^4\theta F(W-\bar W,V+\bar W,\bar V +W)\nonumber\\
&&\qquad -r(V+W)-\bar r(\bar V+\bar 
W)-s(\bar V+W)-\bar s(V+\bar W).
\end{eqnarray}
If we first integrate over the semi-chiral fields, we recover the original 
model in terms of a chiral and a twisted-chiral field. Indeed, 
the equations of motion for $r$ and $s$,
\begin{eqnarray}
D_\p(V+W)=D_+(\bar V+\bar W)=D_-(\bar V+W)=D_\m(V+\bar W),
\end{eqnarray}
imply that
\begin{eqnarray}
D_\p D_-(V-\bar V)=D_+D_\m(V-\bar V)=D_\p D_\m(W-\bar W)=D_+D_-(W-\bar 
W)=0,
\end{eqnarray}
which are solved by putting $V=z$ and $W=w$. The dual model is obtained by first 
integrating over the pre-potentials which yields
\begin{eqnarray}
F_V=r+\bar s,\qquad F_W=r+s,
\end{eqnarray}
which can be solved for the prepotentials $V(r,\bar r, s,\bar s)$ and 
$W(r,\bar r,s,\bar s)$. The semichiral potential,
$K(r-\bar r,r+\bar s, \bar r +s)$, is simply the Legendre 
transform of $F$,
\begin{eqnarray}
&&K(r-\bar r,r+\bar s, \bar r +s)=
F(W-\bar W,V+\bar W,\bar V +W)\nonumber\\
&&\qquad -(W-\bar W)(r-\bar r)-(V+\bar W)(r+\bar s) -(\bar V+W)(\bar r+s).
\end{eqnarray}

In \cite{GHR} it was shown that a system with one chiral and one twisted-chiral superfield 
allows for a full $N=(4,4)$  supersymmetry, 
\begin{eqnarray}
\delta z&=&\eta^\p D_\p\bar w+\eta^\m D_\m w,\nonumber\\
\delta w&=&-\eta^\p D_\p \bar z -\eta^-D_-z,\nonumber\\
\delta\bar z&=&\eta^+D_+w+\eta^-D_-\bar w,\nonumber\\
\delta\bar w &=&-\eta^+D_+ z-\eta^\m D_\m\bar z,
\end{eqnarray}
provided the potential $F(z,\bar z,w,\bar w)$, satisfies the Laplace equation,
\begin{eqnarray}
F_{z\bar z}+F_{w\bar w}=0. \label{lap}
\end{eqnarray}

\vspace{.75cm}


\end{document}